\documentclass{aa}
\newcommand{\dens}[1]{\rho (#1)}
\newcommand{\contr}[1]{\delta (#1)}
\newcommand{\scontr}[2]{\delta_{#1} (#2)}
\newcommand{\contrk}[1]{\delta_{#1}}
\newcommand{\mean}[1]{\left\langle #1 \right\rangle}
\usepackage{psfig}
\begin{document}

   \thesaurus{
              04         
              (11.06.1;  
               11.03.3;  
               02.08.1;  
               03.13.4)} 
   \title{Cooling Limits on Galaxy Formation}

   \subtitle{Gas dynamical simulations incorporating a background UV field
             and metal enrichment}

   \author{Daniel K{\"a}llander \and John Hultman}

   \offprints{J. Hultman}

   \institute{Uppsala Astronomical observatory,
              Box 515,
              S--751 20 ~Uppsala,
              Sweden\\
              e-mail: pohlman@astro.uu.se
              }

   \date{Accepted}

   \maketitle

   \begin{abstract}

We present hydrodynamical simulations of the formation of galaxies
in the mass range $10^9 - 10^{13} {\rm M}_{\odot}$, with the focus on the efficiency
of gas cooling.

The effect of a background UV radiation field, and the
effect of metal enrichment of halo gas due to star formation and stellar
evolution, are investigated.
A background radiation field is found to strongly suppress the formation
of galaxies with circular velocities less than $\approx 50$ km/s.
The effect is, however, not large enough to reconcile hierarchical
clustering models with observations.

Metal enrichment of the halo gas increases the cooling rate at low
redshifts. We find that the mass fraction of gas at virial
temperatures may be reduced by a factor of two, in simulations with a UV 
background field added. The decrease in overall efficiency of gas cooling due
to the inclusion of a UV background field can be more than compensated
for by the increased cooling that follows metal enrichment of halo gas,
but the effect may depend strongly on the assumed model of metal
enrichment.

      \keywords{numerical methods -- hydrodynamics -- gal\-ax\-ies - formation}

   \end{abstract}

\section{Introduction}
It is generally believed that galaxies formed by gravitational collapse.
Small fluctuations in the primeval density field grew through gravitational
instabilities. When the mean density in a proto-galaxy reaches approximately
twice the mean density of the Universe,
the proto-galaxy breaks away from the general expansion of the Universe
and collapses, and forms a gravitationally bound galactic halo.
If the gas in a galactic halo is able to cool efficiently, it can dissipate
kinetic energy and condense into a galactic object at the center of the halo.

The purpose of this paper is to study under what circumstances the gas
in proto-galaxies is able to cool efficiently, and condense into objects
with galactic densities. Particularly, we take a close look at the possible
effects of a strong background UV radiation field and the dynamical effects
of metal enrichment of halo gas. The results are based
on three-dimensional hydrodynamical computer simulations, and a standard CDM
cosmology is assumed throughout. The results should have some relevance
for all hierarchical scenarios of galaxy formation.

In hierarchical models of galaxy formation, the statistical cosmological
time of collapse for a proto-galaxy is an increasing function of mass.
More massive galactic halos thus form later.
This general picture is consistent with observations indicating that clusters
are younger than the contained galaxies, and that the Universe is homogeneous
on very large scales.

Any theory of galaxy formation must address fundamental properties
of the observed galaxy population like characteristic masses and sizes.
Rees \& Ostriker (\cite{Rees77a}), and Silk (\cite{Silk77a}), showed that the
characteristic masses and sizes of galaxies
are just those to be expected of gas clouds that can cool in a Hubble time.
This provides a natural explanation for the sharp upper cut-off in
the galaxy luminosity function; more massive proto-galaxies can not cool
in a Hubble-time.
These calculations were based on simple models of homogeneous gas clouds,
very unlike the highly inhomogeneous collapse of a proto-galaxy in a
hierarchical model, where objects are formed by merging of smaller ones,
where the gas have already had time to cool. If a substantial fraction
of the gas is still heated to virial temperatures, these models still
shows that limits due to gas cooling are likely to play an important
role in the formation of galaxies.

Observations sugg\-est that only a small frac\-tion, a\-round $1-2 \: \%$,
of the closure density of the Universe is in the form of stars.
It therefore seems likely that most
of the baryonic matter has avoided falling into compact objects, where
efficient star formation could take place. 
Hierarchical models (White \& Frenk \cite{White91a}, Cole et al. \cite{Cole94a}),
on the other hand, tend to lock up most of the
baryonic mass in the universe in small galactic objects, at early times.
If the gas in these halos is able to cool and condense into dwarf galaxies,
hierarchical models seem to over-predict the current mass fraction of stars
in the Universe. One possible explanation is that supernova driven winds
are able to suppress
the formation of dwarf galaxies (Dekel \& Silk \cite{Dekel86a}). Another process
that may be important
involves the presence of a photo-ionizing UV background radiation field.
A photo-ionizing background field would heat the gas, and, by keeping the
gas at a high degree of ionization, it would suppress collisional line
cooling from neutral atoms. Evidence for such a field is given by the
lack of a Gunn-Peterson effect even at high redshifts $z\approx 5$
(Webb et al. \cite{Webb92a}), indicating that
the Universe was highly ionized at these redshifts.

Based on simple models of the collapse of gas clouds, Efstathiou 
(\cite{Efstathiou92a}) argued
that a strong background UV radiation field might indeed suppress the formation
of dwarf galaxies. The gas is heated by photo-ionization,
and the gas is also kept at a high degree of ionization that suppresses
collisional line cooling. Efstathiou found that this might prevent the
gas in galactic halos with circular velocities less than $50 km/s$ from cooling.

The UV background photo-ionization heating is able to raise the equilibrium gas
temperature to only about $10^4 - 10^5$ K, for the typical range of densities
in question, ($n_H = 10^{-6} - 10^{-2} \; {\rm cm}^{-3}$). Further, the
photo-ionization
process is proportional to the density, through relative abundances,
but the excitation and collisional cooling processes are proportional to the
density squared. Therefore, the effect of photo-ionization is larger for low
densities, and becomes less significant for higher densities. Above
$10^5-10^6$ K photo-ionization effects also diminishes, due to first hydrogen,
later helium, becoming completely ionized. It is therefore only likely to be
important for the dynamics of small galaxies. But, large galaxies could also
have been affected, since their lower mass progenitors may have been unable
to cool. 

Typical virial temperatures of galactic halos are $10^4-10^6$ K. In this
temperature range the cooling rate of interstellar gas is highly dependent
on the metallicity. The cooling rate of a gas with solar metallicity can
be more than one order of magnitude larger than for a gas of primordial
composition (Sutherland \& Dopita \cite{Sutherland93a}). Observations of intra-cluster gas
typically indicates metallicities of $[{\rm Fe/H}]=-0.4$. This metal enrichment
can have a strong effect on the amount of gas that is able to cool at
low redshifts.

K\"allander (\cite{Daniel96a}) has performed simulations similar to
the ones  presented here, using a similar model for the metal enrichment
of halo gas, (described in the next section), but twice as high metallicity
yield. The numerical resolution was similar, but no UV field was included,
and only galaxies with a total mass of $10^{12} {\rm M}_{\odot}$ were studied.
The increase in cooling rate, due to line cooling in metals, was in that case
found to be enough to cause the hot halo gas to collapse in a cooling flow,
in roughly one Hubble time. Without the increased cooling rate, significantly
less cooling flow was present, and essentially no disk formed.
Some experimentation seems to indicate that the magnitude of the gas
dynamical effects of metal enrichment,
are sensitive to the details of how the metal enrichment is modelled.

White and Frenk (\cite{White91a}) used
an intricate semi-analytic frame\-work to com\-pare diff\-erent mod\-els of hierar\-chical
gal\-axy formation to observations, and found that the results were
sensitive to the assumptions made for the metal enrichment of the gas.

Other simulations includ\-ing the effects of photo-\-ioniza\-tion have been
presented by Vedel, Hellsten, \& Sommer-Larsen (\cite{Vedel94a}),
Thoul \& Weinberg (\cite{Thoul96a}), Quinn, Katz, \& Efstathiou
(\cite{Quinn96a}), Weinberg, Hernquist, \& Katz (\cite{Weinberg97a}),
Navarro \& Steinmetz (\cite{Navarro97a}). These authors uses different
implementations, initial conditions and strengths of the photo-ionizing field.
Vedel et al. uses three dimensional SPH simulations of a Milky Way sized
object, but cannot draw conclusions on objects of other masses.
Thoul \& Weinberg uses one dimensional models, and have much higher resolution. 
The remaining uses 3D SPH simulations and emphasise on resolution effects,
which are shown to be of critical importance to the results in these simulation.
In spite of that these authors uses different implementations, the main
conclusions are very similar. A background photo-ionizing field does have
some effects on lower mass objects, but cannot alone be responsible for
suppressing the formation of dwarf galaxies, and make the hierarchical CDM models
consistent with observations. These authors does not include the effects
of metal enrichment, but this does not alter the general result, however.

Hierarchical clustering is a highly inhomogeneous process. Direct 
three-dimensional simulations are therefore a vital complement to lower
dimensional analytical models. Gas dynamics must be included because
dissipation is a fundamental part of galaxy formation.
In this article we present computer simulations of galaxy formation that
provide further insight on hierarchical clustering
models, with an emphasis on cooling arguments.
Even though a CDM cosmological model is assumed throughout,
the results should have some relevance for all hierarchical scenarios.
The effects of a UV background is investigated by running otherwise identical
simulations, with and without the effects of an ionizing field incorporated.
To be able to address questions about the mass dependence of physical
processes, simulations covering a wide range of galactic masses are
presented.

\section{Simulations}
The simulations presented here follow the evolution of
two fluids: a collision-less component that comprises $95\%$ of the mass
and represents the dark matter, and a second hydrodynamical gas
component. Gravitational forces are calculated using a tree code
and the hydrodynamical equations are solved for the gas by use
of the smooth particle hydrodynamics (SPH) method. Details of the
implementation can be found in Hultman \& K\"allander (\cite{Pohlman97b}).

The gas consists of hydrogen and a $24\%$ mass fraction of helium, typical
of estimated primordial values. Furthermore, we assume that the gas
is optically thin and in ionization equilibrium at all times. 
The most important cooling processes are radiative cooling by bound-bound
and bound-free transitions.
Compton cooling by electron scattering against the Cosmic
Microwave Background (CMB) can also have a significant effect at
high redshifts, but at low redshifts, $z < 4$, the photon density of the
CMB is too low to lead to any significant Compton cooling.

In some of the simulations an external strong UV radiation background field
is present.
The time evolution of this field is fixed already at the start
of the simulations, and does not depend on the state of the gas in the
simulated region.
This field changes the gas cooling rate, and adds a redshift dependence
to the cooling function. The gas is also heated by this field.

Radiative molecular cooling
is not included, and this limits us when choosing objects to study.
Below $\approx 5\times 10^3$ K, atomic radiative cooling is inefficient and
molecular radiative cooling may
be important. By neglecting molecular cooling we cannot with
any certainty simulate systems where the gas pressure at temperatures
below $10^4$ K can be dynamically important.
This implies a lower mass limit on the applicability of these simulations.

Neutral gas of primordial abundance is stable against collapse in a dark
matter halo if the halo mass, $M_{halo}$, is (Rees \cite{Rees86a},
Quinn et al. \cite{Quinn96a}) 
\begin{eqnarray}
M_{halo} & < & 1.09\times 10^9 \left( \frac{T}{10^4 {\rm K}} \right)^{3/2}\times
           \nonumber\\
  & & h_{50}^{-1} (1 + z)^{-3/2} \left( \frac{\mu}{m_p} \right) {\rm M}_{\odot}.
\end{eqnarray}
T, $m_p, \mu, z, h_{50}$ are respectively the gas temperature, the proton mass, 
the gas mean molecular weight, the formation redshift and the Hubble con\-stant
in units of km $\rm s^{-1}$ $\rm Mpc^{-1}$. For a neutral gas, $T \sim 10^4$ K,
and a formation redshift of $z=3$ this corresponds to a mass of
$1.25\times 10^8 {\rm M}_{\odot}$.
A further complication at these mass scales is that galactic halos with as
low masses as $10^8 {\rm M}_{\odot}$ are very susceptible to disruption by supernovae 
(Dekel \& Silk \cite{Dekel86a}). The least total mass in the sequence of
simulations, was for these reasons chosen to have a total mass of
$10^9 {\rm M}_{\odot}$.

Each object was simulated both with and without a photo-ionizing background,
using the same initial density field.
In this way it is possible to investigate the
effects of the photo-ionizing background, without making a statistical study
that would require a large number of simulations.
The assumed background radiation field is of the form
\begin{eqnarray}
J(\nu ) = J_{-21}(z) \times
10^{-21} (\nu_L/\nu) \nonumber\\ {\rm erg \: cm^{-2} sr^{-1} Hz^{-1} s^{-1}},
\end{eqnarray}
where $\nu_L$ is
the Lyman limit frequency, and the time dependence of the field is
contained in the function $J_{-21}(z)$.

Observational determina\-tions of $J_{-21}(z)$ in re\-cent years in\-clude the
follow\-ing (Haardt and Madau \cite{Haardt96a}):
$\log J_{-21}$ $\approx$ $0.5$ at $1.6 < z < 4.1$ (Bechtold \cite{Bechtold94a}),
$-1.0 < \log J_{-21} < -0.5$ at $z\approx 4.2$ (Williger et al.
\cite{Williger94a}),
$\log J_{-21} = 0.0 \pm 0.5$ at $1.7 < z < 3.8$ (Bajtlik et al.
\cite{Bajtlik88a}, Lu et al. \cite{Lu91a}),
and $\log J_{-21} = -0.3 \pm 0.2$ at $1.7 < z < 3.8$ (Espey \cite{Espey93a}).
A decline in the intensity is expected at high redshifts due to a decline
in the number density of quasars and increased absorption by Lyman-alpha
clouds. Here $J_{-21}(z)$ is taken to have the form
\begin{equation}
  J_{-21}(z) =
  \left\{ \begin{array}{l@{\quad \quad}l}
    4/(1 + z) & 3 < z < 6 \\
    1 & 2 < z < 3 \\
    z - 1 & 1 < z < 2. \\
  \end{array} \right .
\end{equation}

\begin{figure}
  \psfig{file=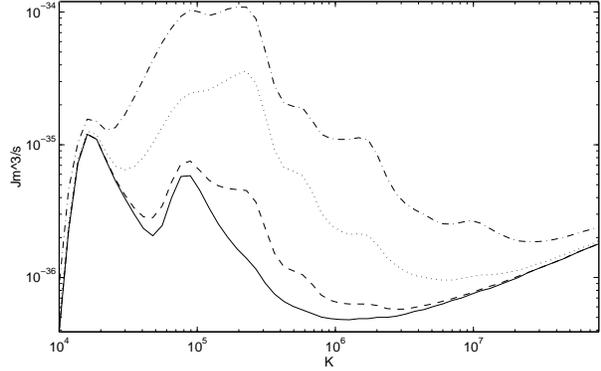,width=8cm}
  \caption{\label{coolfunc} The normalized cooling function as given
  by Sutherland \& Dopita. Different curves correspond to different
  gas metallicity, zero metallicity ([Fe/H] = - 3, solid curve),
  [Fe/H] = - 2 (dashed curve), [Fe/H] = - 1 (dotted curve) and [Fe/H] = 0
  (dot-dashed curve). ([Fe/H] being the value of the logarithm of the
  metal content normalized to the solar value.)}
\end{figure}

Proto-galactic gas is enriched with metals ejected from heavy stars,
through stellar winds and supernovae. This leads to an significant increase in
the gas cooling rate at the temperatures that are relevant for
galaxy formation, as can be seen from cooling curve by Sutherland \& Dopita (\cite{Sutherland93a}), in Fig \ref{coolfunc}.
To take cooling effects of metals into account, a description of the
star formation rate is necessary.
Star formation is, however, not well understood.
Since the only purpose here is to roughly estimate the amount of heavy
metals ejected, a simple description is adequate.

All gas with a density contrast above 200 is taken to be inside "collapsed
structures", and gas inside such collapsed structures is assumed to be
transformed completely into stars. This is only made for the purpose of
estimating the average gas metallicity in the proto-galaxy; dynamically,
baryonic mass remains gaseous everywhere. Instantaneous recycling is also
assumed (heavy stars eject their metals immediately after being formed)
and complete mixing of the metals throughout the proto-galaxy (making the
gas metallicity homogeneous) which simplifies the model further. The gas
metallicity is then given by (Tinsley \cite{Tinsley80a})

\begin{equation}
Z(t) = y \ln{M/(M - M_s(t))}
\end{equation}
where $Z$, $y$, $M$, $M_s$ and t are, respectively, gas metallicity, yield, total baryonic mass, total mass in stars and time. That is, the average gas
metallicity in the proto-galaxy, is only dependent of the average gas
fraction in the proto-galaxy, (the proto-galaxy being the entire simulation
volume in this context). Here a yield $y=0.5 {\rm Z}_{\odot}$ is
adopted, which was found to give a gas metallicity in accordance with
observed values.

Assuming a time evolution and spectral shape for the background field,
a four dimens\-ional table in redshift, met\-all\-icity, density and temperature
was calculated for both the cooling and heating. 
We used the publicly available program CLOUDY (Ferland \cite{CLOUDY})
to calculate cooling and heating tables.
The tables were stored in a file and subsequently read in at the
beginning of each simulation.
Linear interpolation was then used to calculate heating and cooling rates
during the simulation.


\section{Initial conditions}

In setting up initial conditions for a simulation it is necessary to
specify the cosmological model. Here a CDM model, with a baryon fraction
of $5\%$, a bias parameter of $b=1.5$, $\Omega = 1$ , $\Lambda = 0$ ,
and a Hubble constant $h=0.5$ in units of 100 km/s/Mpc, is employed.
The main significance of the cosmological model is in this case to provide
a normalization, and shape, of the power spectrum of density fluctuations.
The amplitude of the power spectrum on galactic scales is constrained by
observations, choosing $\sigma_8 = 1/b$, (the rms field value, smoothed with
a top hat filter on a scale of $8/h$ Mpc). Choosing another hierarchical
cosmological model is therefore likely to have only moderate effects on the
results of these simulations.

The simulations presented here cover a mass range of objects,
$10^{9} - 10^{13} {\rm M}_{\odot}$, in order to address questions about the mass
dependence of different physical mechanisms.
The gas is initially represented by $8000$ particles
and the dark matter by $4000$ particles. Gas particles are under certain 
conditions allowed to merge with other nearby gas particles, and the
number of gas particles is therefore a decreasing function of time,
(see Hultman \& K\"allander \cite{Pohlman97b} for details).
The simulations start with a spherical region of the Universe at a
redshift of $z = 30$.

We choose the simulation volumes as spheres, containing precisely the mass
of the object in question. This minimizes the volume needed to be
simulated, reducing costs in computational time, allowing for highest
possible resolution, which is of critical importance in these type of
simulations. The initial conditions used here are similar to the ones used
by Katz \& Gunn (\cite{Katz91a}), but there are some important differences,
and improvements as will be explained below.

Selecting a limited spherical region
to study the formation of a galaxy leads to two major complications,
the neglect of the surroundings and the selection of a proper site.

Galaxies do not form at random locations. Finding good candidates
for galactic seeds at high redshifts, using only the density distribution as
a guide, is a difficult task. One possible procedure is
to make a low resolution simulation of a large region, and thereby identify
the sites where galactic halos of the desired type form. After that, such
a region can be re-simulated with higher resolution. Although cumbersome,
this is feasible and has been done (Navarro \& White \cite{Navarro94b}).

We chose instead to rely on the ``peaks formalism'' (Bardeen et al.
\cite{BBKS86a}) to identify likely sites of the formation of galactic halos.
This enables us to use an explicit scheme where key properties of the
proto-galaxy can be specified, ``seeding'' the formation of the galactic
object to form at the desired location. The details of this are described in
Appendix \ref{app_a}.

By ignoring the surrounding regions outside the proto-galaxy, no account
is taken of tidal interactions. Tidal interactions with the surrounding
matter is believed to be what gives galaxies their spin.
In its early pre-collapse phase, the proto-galaxy acquires angular momentum
thro\-ugh gravitational interactions with surrounding matter.
Density fluctuations grow by gravitational instabilities and at
some time the proto-galaxy collapses.
When it
collapses, the quadrupole, and higher, moments of its mass distribution
diminish as it shrinks in size. This drastically reduces any tidal
interactions, and the angular momentum accumulated in the pre-collapse phase is
effectively frozen in at this time.
It is therefore believed that a galaxy that has not been involved in any
major merger events has had an almost constant angular momentum since the
main collapse of the proto-galaxy.


Thus, to approximate the neglected effects of the proto-\-galactic surroundings,
the simulated spherical region is started in solid body rotation. The
amount of angular momentum added corresponds to a spin parameter of $\lambda = 0.05$ (Barnes \& Efstathiou \cite{Barnes87a}).

\begin{table}
\caption[]{\label{startpars} Simulation parameters.}
\begin{flushleft}
\begin{tabular}{cllllllllllllll}
\hline\noalign{\smallskip}
Total mass & $r_{init}$ & $V_{circ}$ & $\epsilon_g$ & $\epsilon_{DM}$ & $T_{vir}$ \\
\noalign{\smallskip}
[${\rm M}_{\odot}$] & [Mpc] & [km/s] & [kpc] & [kpc] & [K]\\
\noalign{\smallskip}
\hline\noalign{\smallskip}
$10^{9}$ & 0.15 & 28 & 0.37 & 1.0 & $2.85\cdot 10^4$ \\
$10^{10}$ & 0.33 & 55 & 0.74 & 2.0 & $1.10\cdot 10^5$ \\
$10^{11}$ & 0.71 & 100 & 1.1 & 3.0 & $3.63\cdot 10^5$ \\
$10^{12}$ & 1.52 & 170 & 1.5 & 4.0 & $1.05\cdot 10^6$ \\
$10^{13}$ & 3.27 & 330 & 2.2 & 6.0 & $3.95\cdot 10^6$ \\
\noalign{\smallskip}
\hline
\end{tabular}
\end{flushleft}
\end{table}

\section{Primordial gas simulations}

A first series of simulations does not include a UV field, nor metal
enrichment of gas. The gas has a primordial composition, with a 24\%
mass fraction of helium, and the rest of the mass in hydrogen.

\begin{figure}
  \psfig{file=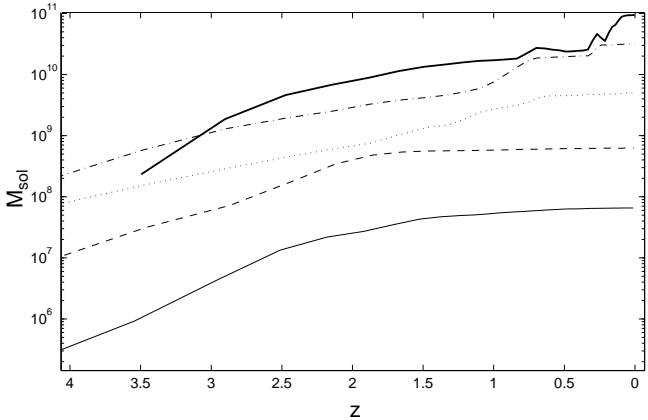,width=8.8cm}
  \caption{ \label{mainobjmassplain} The mass of the most massive progenitor,
as a function of redshift, for the primordial gas simulations.
The different curves represent values for
simulations with different total mass: $10^{9} {\rm M}_{\odot}$ (solid), $10^{10} {\rm M}_{\odot}$ (dashed), $10^{11} {\rm M}_{\odot}$ (dotted), $10^{12} {\rm M}_{\odot}$ (dot-dashed), $10^{13} {\rm M}_{\odot}$ (fat solid).
}
\end{figure}

All the simulations in this series result in the formation of a single
dominant collapsed gas object at $z=0$, which contain more than 90\% of
the collapsed gas mass. This object is built up through hierarchical merging
of smaller objects. The mass of the most massive progenitor as a function
of redshift can be seen in Figure \ref{mainobjmassplain}.
The redshift, at which the mass of the largest progenitor
has acquired half of the final mass, is an increasing function of total
proto-galactic mass. For the $10^9 {\rm M}_{\odot}$ simulation this redshift is
$z \approx 2.2$, and for the  $10^{13} {\rm M}_{\odot}$ simulation it is
$z \approx 0.2$.

\begin{figure}
  \psfig{file=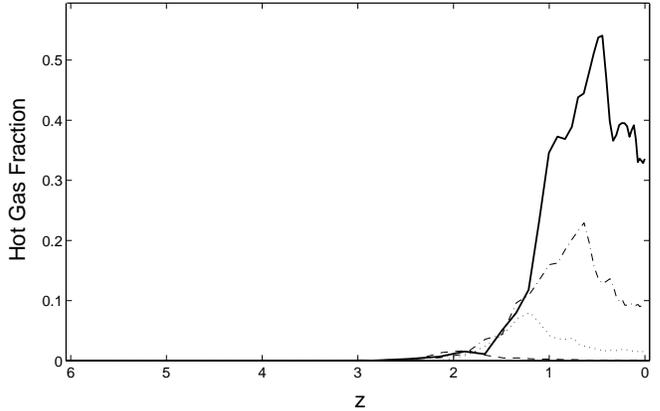,width=8.8cm}
  \caption{ \label{hotvirfracplain} The mass fraction of gas inside the
virial radius that has a temperature exceeding half the virial temperature,
for the primordial gas simulations.
Notation as in Figure \ref{mainobjmassplain}.
}
\end{figure}

During the collapse of the proto-galaxy the gas acquires kinetic energy,
which is then converted into thermal energy by shocks and radiated away.
In the low mass galaxies the radiative gas cooling is efficient and
keeps the gas cool, as can be seen in Figure \ref{hotvirfracplain}.
The fraction of gas that can cool depends on the redshift of formation.
At high redshifts the density of the Universe is higher, and the cooling 
therefore stronger.


The shape of the cooling curve
also gives a strong implicit redshift dependence of the efficiency of gas
cooling. Radiative energy losses are highest in the temperature range
$10^4 - 10^5$ K. Most gas that is shock heated to temperatures above $10^5$ K
stays hot for more than a Hubble time, whereas gas that is
heated to less than $10^5$ K experiences an order of
magnitude stronger cooling and cools well within a Hubble time.

It can be seen in Figure \ref{hotvirfracplain} and Table \ref{simpars} that the galactic halos
that contain gas at virial temperatures at $z=0$, are those with a virial
temperature exceeding $10^5$ K. These galactic halos form late, and have
accumulated less than half the final mass at $z=1$.

The gas core in the $10^{13} {\rm M}_{\odot}$ simulation forms late,
at $z\approx 0.2$, see Figure \ref{mainobjmassplain}, in a series of
merging events. However, the hot halo
remaining at $z=0$, and containing $30\%$ of the gas mass, forms much
earlier, at $z\approx 1$. Between $z=1$ and $z=0.2$, several collapsed gas
clumps, none of which has a mass in excess of 20\% of the mass of the
final gas core, share a common hot pressure supported gas halo.
The gas cooling is relatively inefficient at the high temperatures of
this hot halo, and there is no significant amount of cooling inflow
of gas until the present epoch.

\begin{figure}
  \psfig{file=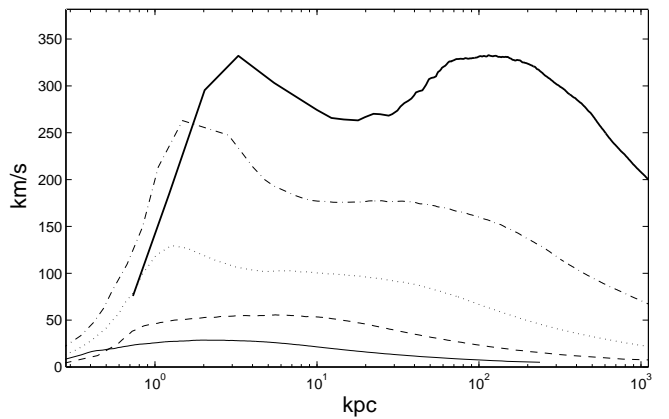,width=8.8cm}
  \caption{ \label{rotcurveplain} Circular velocity as a function of radius,
for the primordial gas simulations.
Notation as in Figure \ref{mainobjmassplain}.
}
\end{figure}

The central gas core that forms in all simulations is very compact, and
is not resolved.
This over-concentration of mass can be seen from the rotation curves
in Figure \ref{rotcurveplain}, which rise very rapidly in the innermost
regions.

The galactic objects that form are very compact
because the collapsed gas has lost much of the angular momentum that
was injected at the start of the simulations. This effect is well known, and
have been previously investigated by Navarro \& Benz (\cite{Navarro91a}),
Katz \& Gunn (\cite{Katz91a}), Vedel, Hellsten, \& Sommer-Larsen
(\cite{Vedel94a}), Navarro, Frenk, \& White (\cite{Navarro95a}),
Navarro \& Steinmetz (\cite{Navarro97a}), but was first found by
Lake \& Carlberg (\cite{Lake88a}), (using a ``sticky particle'' method).

\begin{figure}
  \psfig{file=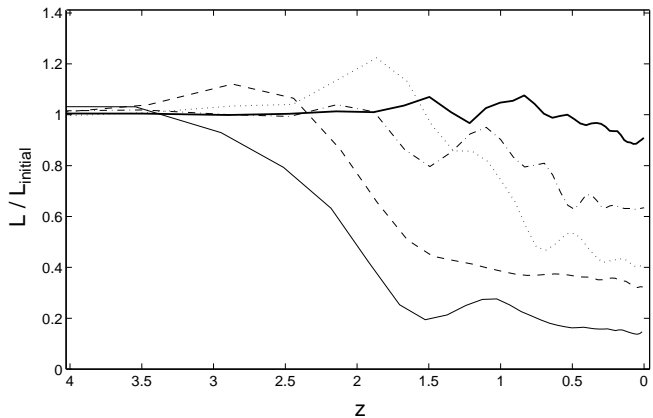,width=8.8cm}
  \caption{\label{angmomplain} Total angular momentum of the gas component,
normalized to the initial value, as a function of redshift, for the primordial 
gas simulations.
Notation as in Figure \ref{mainobjmassplain}.
}
\end{figure}

The total angular momentum
is in all cases conserved to well within 1\%, and if the collapsed gas has
lost angular momentum that angular momentum must have been transferred
to the dark matter component. Figure \ref{angmomplain} shows the total
angular momentum of the gas, normalized to the initial value. It is clear
that between 10 and  90\% of the gas angular momentum has been
transferred to the dark matter component. In the $10^{12}$ and
$10^{13} {\rm M}_{\odot}$ simulations this angular momentum transfer is, however,
not very
pronounced. Instead, most of the gas angular momentum is contained in
the hot halo of pressure supported gas that surrounds the central collapsed
object. (Note that above mentioned authors considered the angular momentum
of the disk, and did not include the halo.)

\section{Simulations with an ionizing field}

A background UV field affects the gas in two ways: (i) the gas
is heated by photo-ionization; and (ii) it ionizes the gas and thereby
reduces its ability to cool by collisional excitation of
neutral atoms. In these simulations the background radiation field
is non-zero in the range $1<z<6$.


The rate of photo-ionization heating by the back\-gro\-und field is
proportional to the local gas density, and the rate of cooling by
collisional line radiation, the dominant cooling mechanism, is proportional
to the square of the local gas density. Heating by photo-ionization is therefore
most important in low density regions, where the equilibrium temperature
can rise to $T \approx 3\cdot 10^4$ K. This temperature is comparable to
the virial temperature of the galactic halos that form in the $10^9 {\rm M}_{\odot}$
and $10^{10} {\rm M}_{\odot}$ simulations, but much less than the corresponding 
temperature for the largest simulations. The dynamical effects of heating
should therefore be small for the largest galaxies.

\begin{figure}
  \psfig{file=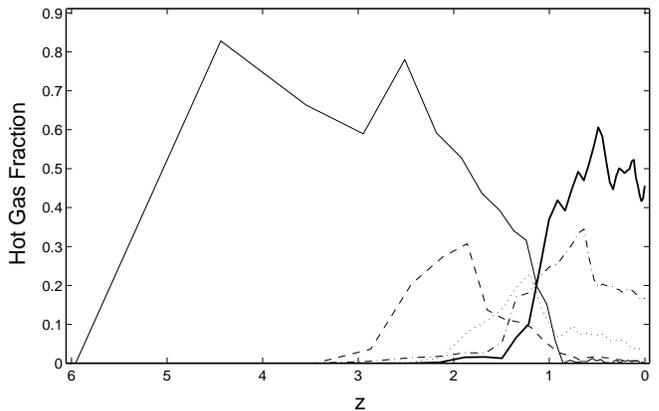,width=8.8cm}
  \caption{ \label{hotvirfracion} The mass fraction of gas inside the
virial radius that has a temperature exceeding half the virial temperature,
for the simulations with a background UV field.
The curves represent simulations including a background radiation field.
Notation as in Figure \ref{mainobjmassplain}.}
\end{figure}

Figure \ref{hotvirfracion} shows the mass fraction of gas within the
virial radius that has a temperature exceeding half the virial temperature
as a function of redshift. The difference is modest for the more massive
galaxies, when comparing with the corresponding curves for the simulations
without a background field. In the $10^{12} {\rm M}_{\odot}$ simulation the
fraction of hot gas at $z=0$ increases from around 10\% to 20\%,
when the background radiation field is included.
On the other hand, the
smallest galaxies show dramatic changes.
In the $10^9 {\rm M}_{\odot}$ simulation, most of the gas is heated to temperatures
above the virial temperature by photo-ionization. This prevents the gas from
falling into the galactic halo potential well. When the background field
falls in strength, after $z = 2$, about $10\%$ of the gas is able to cool and
condense into a galactic object.

\begin{figure}
  \psfig{file=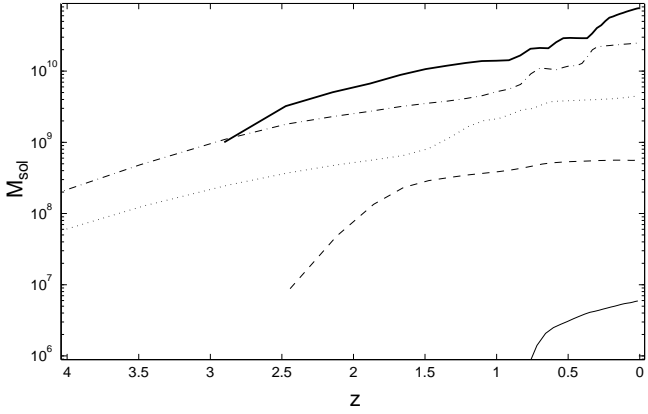,width=8.8cm}
  \caption{ \label{mainobjmassion} The mass of the most massive progenitor,
as a function of redshift, for the simulations with a background UV field.
Notation as in Figure \ref{mainobjmassplain}.
}
\end{figure}

The reduced cooling rate in this series of simulations, with a background
UV field, also reduces the gas mass of the most massive objects that form.
As can be seen when comparing Figure \ref{mainobjmassion} with Figure
\ref{mainobjmassplain}, the reduction in object masses is most pronounced
in the smallest simulations, with a total mass of $10^9 {\rm M}_{\odot}$ and 
$10^{10} {\rm M}_{\odot}$.


\begin{figure}
  \psfig{file=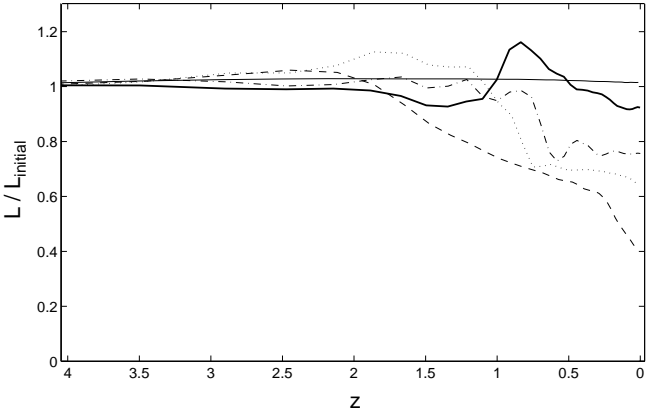,width=8.8cm}
  \caption{ \label{angmomion} Total angular momentum of the gas component,
normalized to the initial value, as a function of redshift, for the simulations
 with a background UV field.
Notation as in Figure \ref{mainobjmassplain}.
}
\end{figure}

There is less mass in the clumpy and cold gas component when a background
UV field is incorporated. This lowers the angular momentum transfer to the dark matter component, as can be seen in Figure \ref{angmomion}, to be compared with
Figure \ref{angmomplain}.

\begin{figure}
  \psfig{file=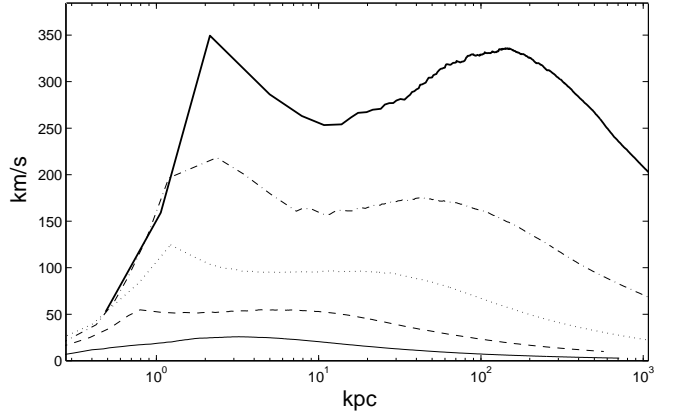,width=8.8cm}
  \caption{ \label{rotcurveion} Circular velocity as a function of radius,
for the simulations with a background UV field.
Notation as in Figure \ref{mainobjmassplain}.
}
\end{figure}

In all cases, the cold collapsed gas is concentrated in a compact
object with an extent of less than a few smoothing lengths. Around this
object is a
second gas component in the form of a hot pressure supported halo. 
In the simulations where
most of the gas can cool and condense into an object, most of the angular
momentum is transferred to the dark matter component. In the cases where
cooling
is less efficient, more angular momentum is retained in the gas component, but
instead the gas angular momentum is contained in the hot gas halo.
The resulting gas cores are still very compact, as can be seen in Figure
\ref{rotcurveion}.

\section{Simulations with an ionizing field and metal enrichment}

The cooling rate of interstellar gas depends sensitively on the composition
assumed. Metals can increase the cooling rate by an order or two of magnitude
in the temperature range relevant here. This increase is mostly due to
collisional line cooling by oxygen and carbon. We employ a very simplistic
scheme to estimate the metallicity of the gas, as described earlier in this 
paper.
We have included the effects of this modulation of the gas cooling rate,
in addition to the inclusion of a background radiation field. To some
extent these two processes are expected to counteract each other, since
a background field limits the ability of the gas to cool, and a high
metal content increases the cooling rate. However, the effects of a
background radiation field vanishes at low redshifts, whereas
the metal content of the gas is only significant at lower redshifts,
after significant star formation has taken place.

\begin{figure}
  \psfig{file=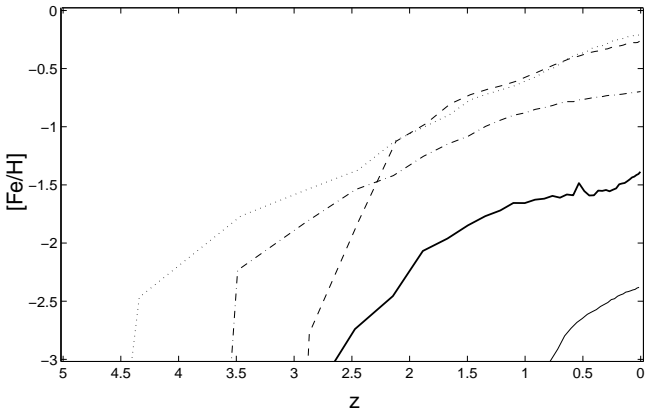,width=8.8cm}
  \caption{ \label{metallicity} Metallicity
as a function of redshift. The values are the logarithm of the metal
content normalized to the solar value.
Notation as in Figure \ref{mainobjmassplain}.
}
\end{figure}

Through the amount of collapsed 
gas it is possible to get a very rough estimate of the star formation rate,
and, with the further assumption of instant mixing, consequently an estimate of the
metal enrichment. Metallicity as a function of redshift is displayed in Figure
\ref{metallicity}, for the different simulations. The gas in the
$10^{9} {\rm M}_{\odot}$ simulation does not reach metallicities high
enough to significantly affect the gas cooling rate, and the result for
this mass is therefore
practically identical to that of the corresponding simulation without metal
enrichment.

\begin{figure}
  \psfig{file=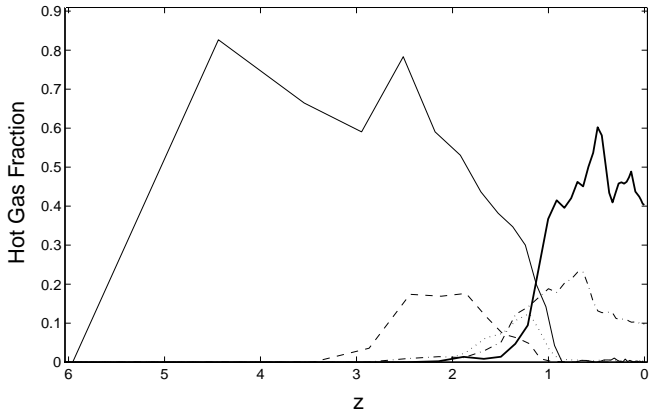,width=8.8cm}
  \caption{ \label{hotvirfracmet} The mass fraction of gas inside the
virial radius that has a temperature exceeding half the virial temperature,
for the simulations with a background UV field and metal enrichment.
These curves are for the simulations including a background radiation field,
and metal enriched gas.
Notation as in Figure \ref{mainobjmassplain}.
}
\end{figure}

\begin{figure}
  \psfig{file=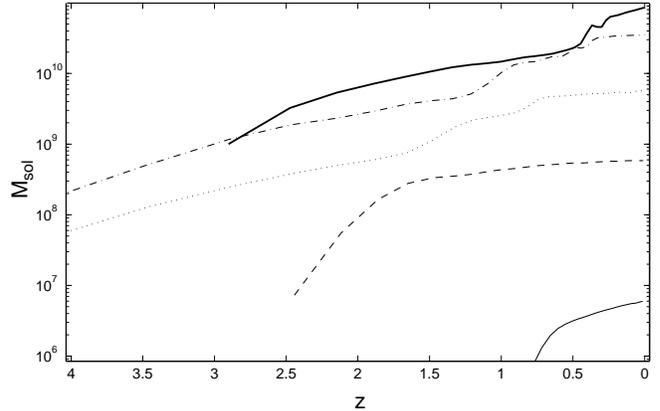,width=8.8cm}
  \caption{ \label{mainobjmassmet} The mass of the most massive progenitor,
as a function of redshift, for the simulations with a background UV field and
 metal enrichment.
Notation as in Figure \ref{mainobjmassplain}.
}
\end{figure}

The early time evolution, for $z>2$, is almost unchanged for all simulated
proto-galaxies, when comparing with the simulations that include a
background field, but no metal enrichment. At later times, the metallicity
of the gas leads to more efficient cooling, as can be seen when comparing
Figure \ref{hotvirfracmet} with Figure \ref{hotvirfracion}.
A hot halo of gas, at $z=0$, is only present in the
$10^{12} {\rm M}_{\odot}$ and $10^{13} {\rm M}_{\odot}$ simulations, and, in fact, all
galactic halos contain less gas at virial temperatures than in the
corresponding simulations without a background field and metal enrichment.
The gas mass of the most massive progenitor is displayed in Figure
\ref{mainobjmassmet}.

\begin{figure}
  \psfig{file=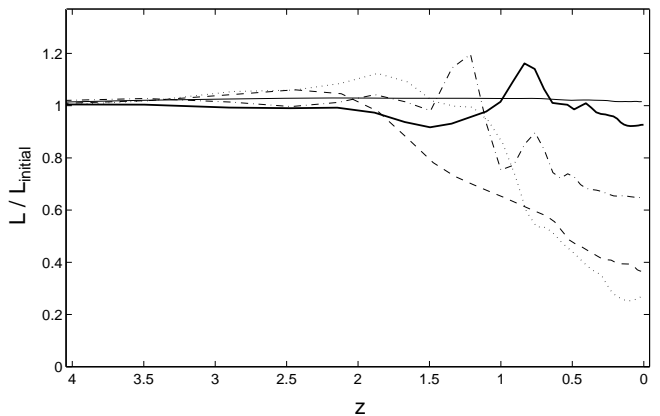,width=8.8cm}
  \caption{ \label{angmommet} Total angular momentum of the gas component,
normalized to the initial value, as a function of redshift, for the
simulations with a background UV field and metal enrichment.
Notation as in Figure \ref{mainobjmassplain}.
}
\end{figure}

\begin{figure}
  \psfig{file=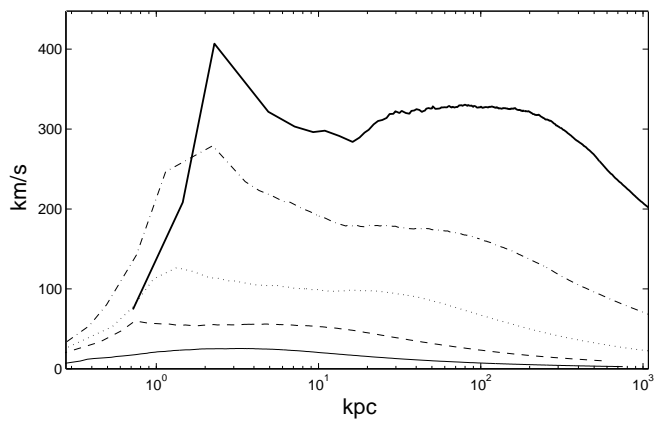,width=8.8cm}
  \caption{ \label{rotcurvemet} Circular velocity as a function of radius,
for the simulations with a background UV field and metal enrichment.
Notation as in Figure \ref{mainobjmassplain}.
}
\end{figure}

\section{Conclusions}

\begin{figure}
  \psfig{file=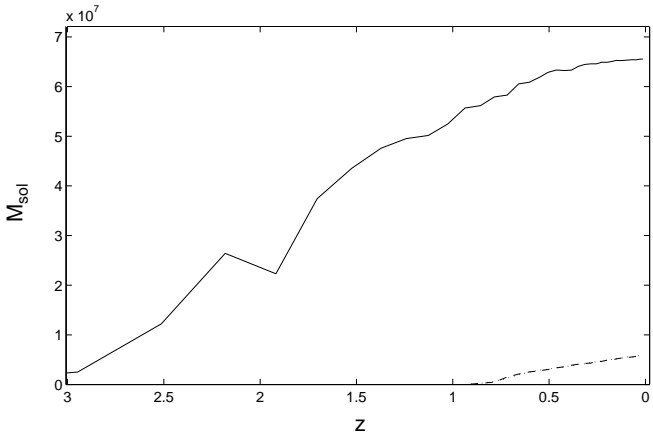,width=8.8cm}
  \caption{ \label{objmass09} Mass of the most massive progenitor, as a
function of redshift.
The three different curves represent values for the $10^{9} {\rm M}_{\odot}$
simulation, evolved in time with each one of the three physical models 
employed primordial gas (solid), primordial gas and background radiation
field (dashed), metal enriched gas and background radiation field (dotted).
On this scale, the curves for the simulations including a background UV
field, with and without metal enrichment, coincide.
Note that the curves represent the collapsed gas mass, the mass of
the dark matter halo is not included.
}
\end{figure}

\begin{figure}
  \psfig{file=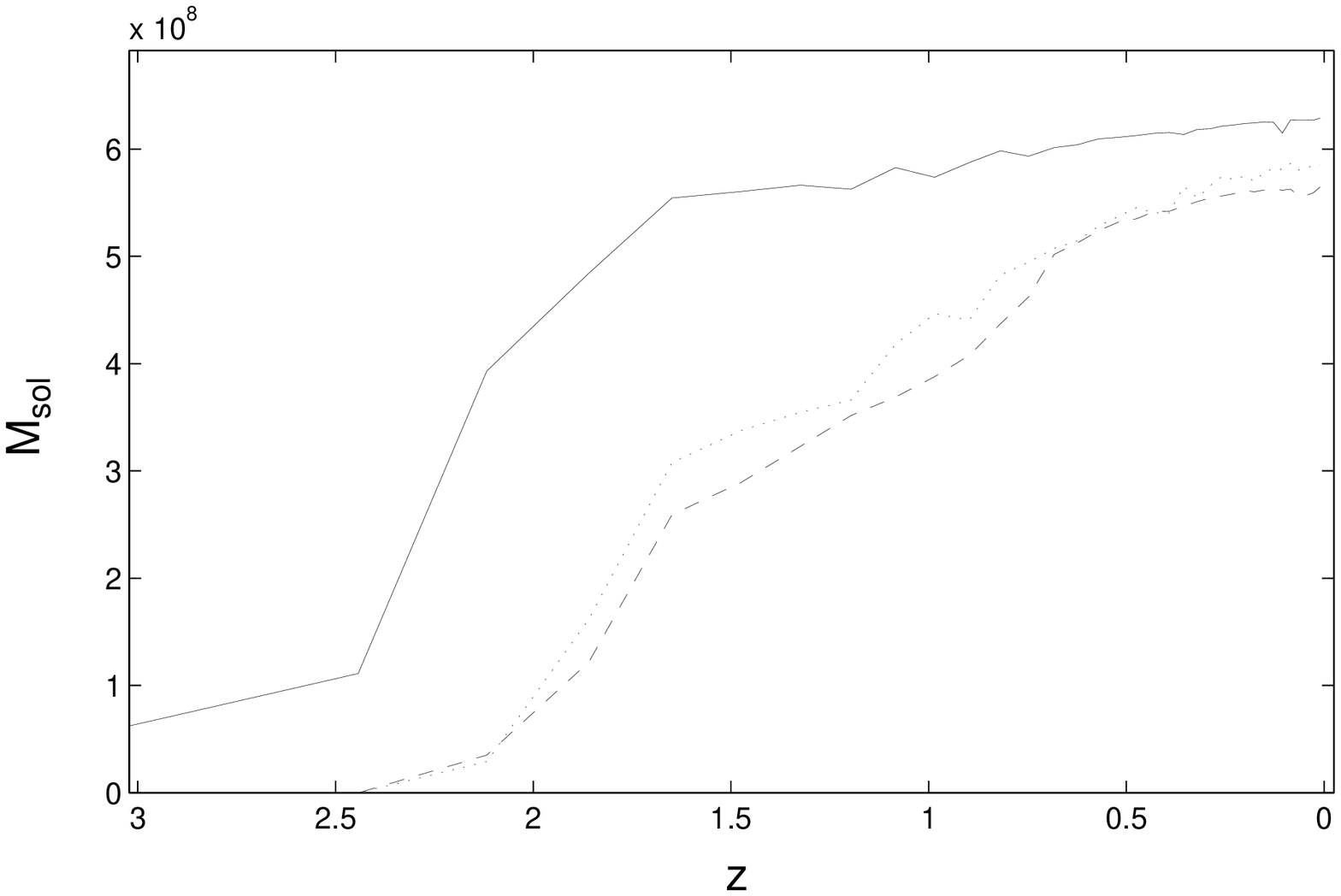,width=8.8cm}
  \caption{ \label{objmass10} Mass of the most massive progenitor, as a
function of redshift.
The three different curves represent values for the $10^{10} {\rm M}_{\odot}$
simulation.
Notation as in Figure \ref{objmass09}.
Note that the curves represent the collapsed gas mass, the mass of
the dark matter halo is not included.}
\end{figure}

\begin{figure}
  \psfig{file=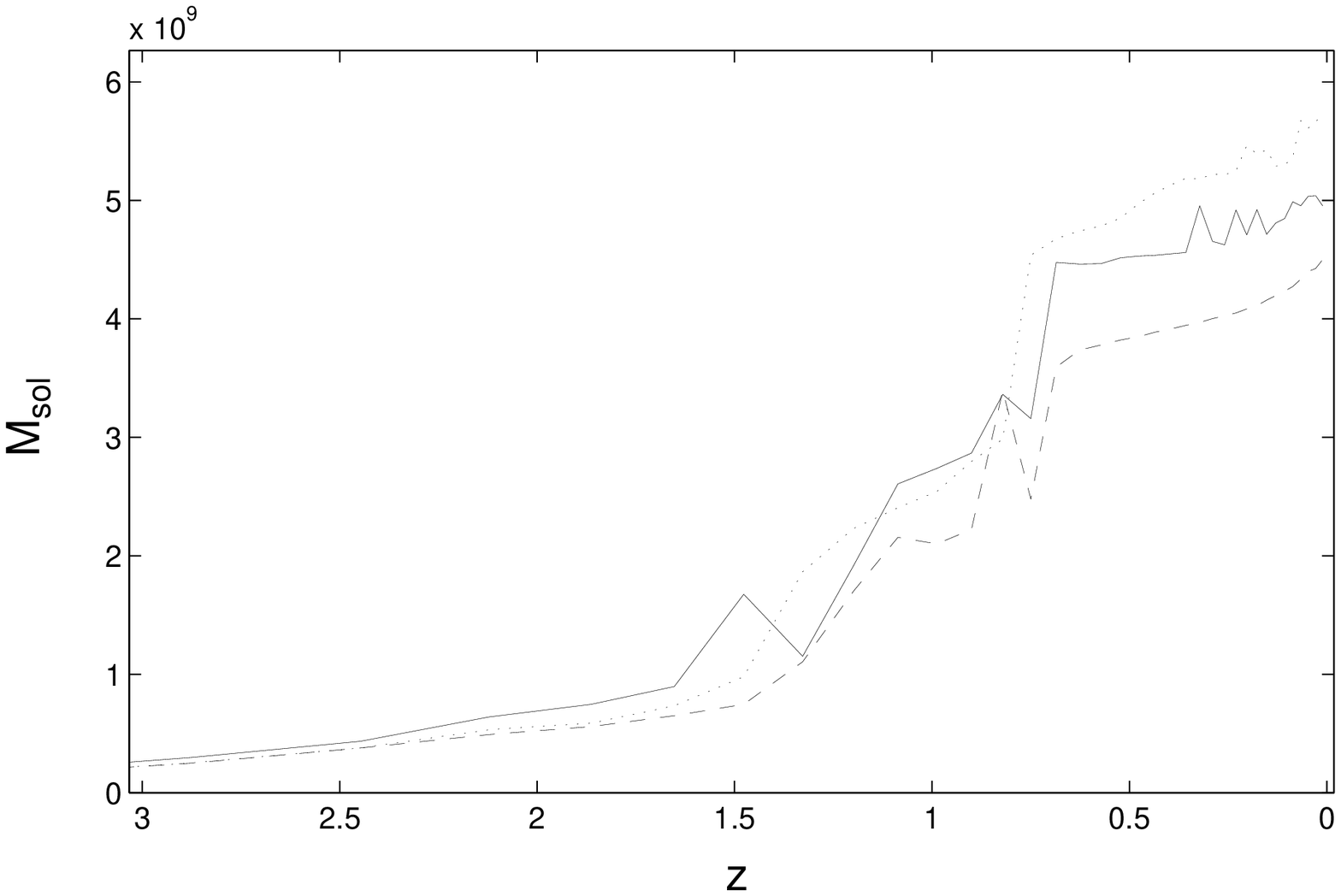,width=8.8cm}
  \caption{ \label{objmass11} Mass of the most massive progenitor, as a
function of redshift.
The three different curves represent values for the $10^{11} {\rm M}_{\odot}$
simulation.
Notation as in Figure \ref{objmass09}.
Note that the curves represent the collapsed gas mass, the mass of
the dark matter halo is not included.}
\end{figure}

\begin{figure}
  \psfig{file=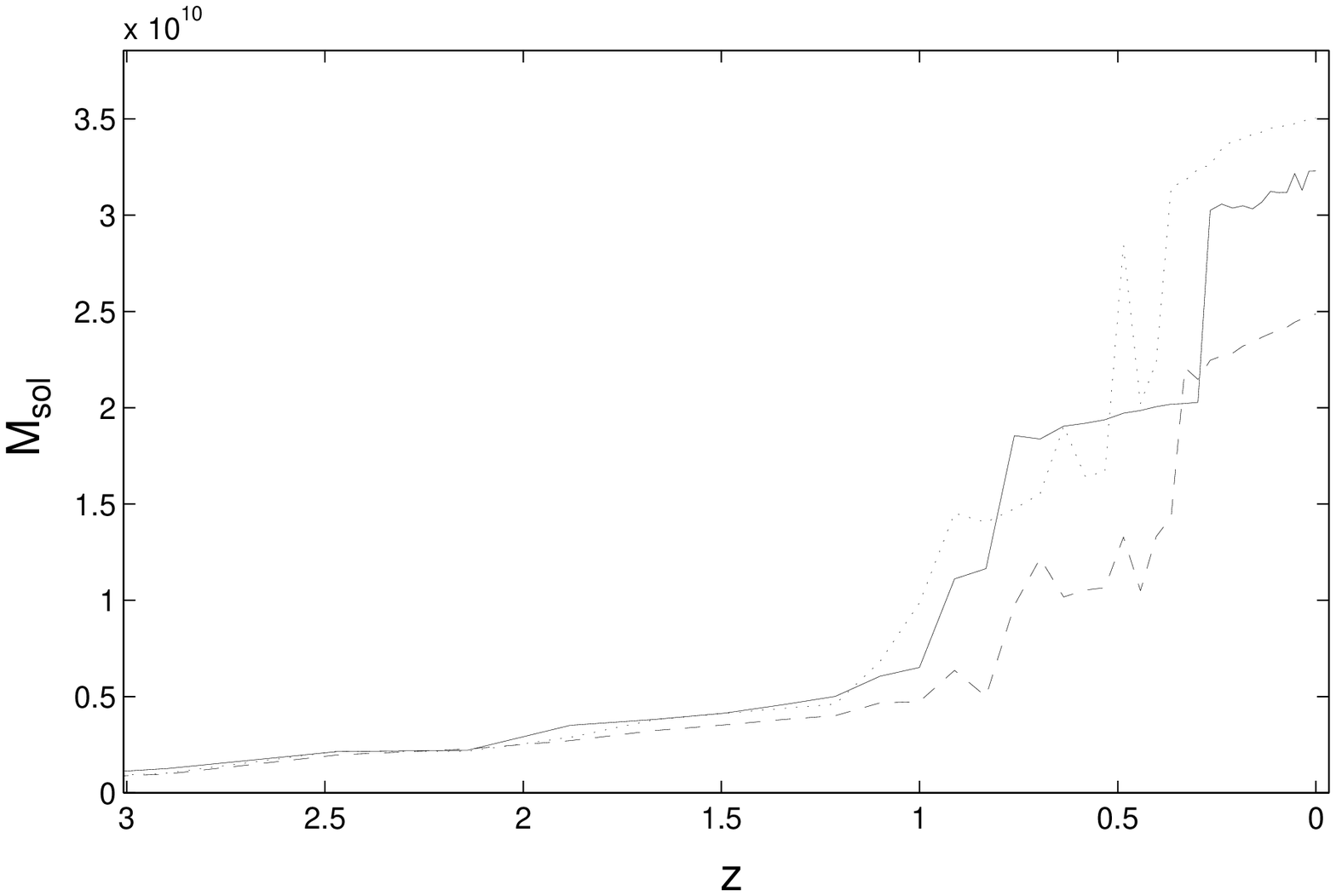,width=8.8cm}
  \caption{ \label{objmass12} Mass of the most massive progenitor, as a
function of redshift.
The three different curves represent values for the $10^{12} {\rm M}_{\odot}$
simulation.
Notation as in Figure \ref{objmass09}.
Note that the curves represent the collapsed gas mass, the mass of
the dark matter halo is not included.}
\end{figure}

\begin{figure}
  \psfig{file=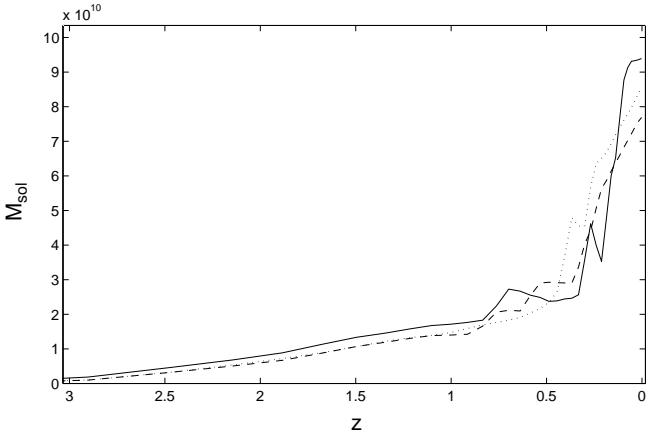,width=8.8cm}
  \caption{ \label{objmass13} Mass of the most massive progenitor, as a
function of redshift.
The three different curves represent values for the $10^{13} {\rm M}_{\odot}$
simulation.
Notation as in Figure \ref{objmass09}.
Note that the curves represent the collapsed gas mass, the mass of
the dark matter halo is not included.}
\end{figure}

The mass of the most massive progenitor as a function of redshift, for
all the simulations presented here, is
shown in Figures \ref{objmass09}, \ref{objmass10}, \ref{objmass11},
\ref{objmass12}, and \ref{objmass13}.

These simulations clarify some points of previous arguments
about galaxy formation that are based on simple analytic models,
and estimates of the efficiency of gas cooling.
It is clear that the inclusion of a background radiation field, consistent
with the observed Gunn-Petersson effect, can strongly suppress the formation
of galaxies with total mass less than $\approx 10^9 {\rm M}_{\odot}$ (circular
velocity $<30$ km/s). Furthermore, the formation of
galaxies with total mass less than $\approx 10^{10} {\rm M}_{\odot}$ (circular 
velocity $<60$ km/s) may be significantly delayed.
These results of course depend on the assumed temporal and spectral form
of the background radiation field. Nevertheless, the results are in
reasonable agreement with previous analytic estimates (Efstathiou
\cite{Efstathiou92a}),
and in good agreement with recently published results based on simulations 
(Quinn et al. \cite{Quinn96a}, Navarro \& Steinmetz \cite{Navarro97a}, 
Weinberg et al. \cite{Weinberg97a}).

Hierarchical models of galaxy formation tend to over-produce galaxies with
circular velocities less than 100 km/s.  Our results indicate that photo-ionization
alone is not sufficient to suppress the formation of these galaxies, since
the effects on galaxies with circular velocities larger than $\approx$ 60
km/s is very limited.

The galactic objects that form in three-dimensional hydrodynamical
simulations, are too compact when compared with observed disk galaxies.
The reason for this is that most of the angular momentum in the gas
component is transferred to the dark matter. Navarro \& Steinmetz
(\cite{Navarro97a}) find that collapsed objects acquire even less
angular momentum, when the effects of a UV field is included. Comparing
Figure \ref{angmomplain} and Figure \ref{angmomion}, these simulations
show that the angular momentum transfer, from the gas to the dark matter,
decreases in magnitude when a UV field is included. This is not a
contradiction. When a UV field is included, most of the gas angular
momentum at $z=0$ is contained in a hot pressure supported halo. The angular
momentum content of the collapsed gas cores does in fact decrease also
in our simulations,
in agreement with the results of Navarro \& Steinmetz (\cite{Navarro97a}).

Metal enrichment of the interstellar gas increases the gas cooling rate
at late times,
and may have significant effects on the amount of gas that may cool and
sink to the center of a galactic halo in a Hubble time.
The inclusion of a background radiation field leads to more massive
hot halos in large galaxies, $10^{12}$ and $10^{13} {\rm M}_{\odot}$, as seen
when comparing Figure \ref{hotvirfracion}
with Figure \ref{hotvirfracplain}. These hot halos contain most of the gas 
angular momentum. Metal enrichment
increases the gas cooling rate at late times, and could potentially
lead to the collapse of the hot halo gas, to the center of the galactic
dark matter halo, in a cooling flow.
However, as seen in Figure \ref{hotvirfracmet}, the increase in cooling
rate is not enough for this to happen. See however Appendix \ref{two_body} for
potential two body heating effects. Increased cooling due to metal
enrichment does decrease the mass of the hot halo that forms,
(Figure \ref{hotvirfracmet}), but most of the
gas angular momentum still resides in the remaining hot halo.

If inhomogeneities in the gas are smoothed out by the limited resolution
used, the average cooling rate in the region will change.
This is a problem common to all hydrodynamical simulations of galaxy formation.
Some implicit assumption must be made, e.g., that the density field is smooth
on unresolved scales due to physical processes not incorporated into the
simulation or, that the density in regions where this could have an effect
is already so high that the cooling is extremely efficient both with and
without unresolved density fluctuations. Previous simulations {\em without} UV
background ionization and heating, did {\em not} suffer from resolution
effects, as badly as one might (naively) expect from the squared density
dependence of the cooling function. The reason being that the cooling function
is divided by the density, i.e. the gas cooling rate, per unit mass,
is (roughly) proportional to the density, when thermal energy is integrated.
For example, Navarro \& White (\cite{Navarro94b}) varied the gas mass-
resolution by a factor of two, and Hultman \& K\"allander (\cite{Pohlman97b}),
by a factor of ten, both showing comparatively small effects.

Then, when a UV background field is included, there is a competition between
density vs. density squared dependencies, being explicitly sensitive to
resolution. Indeed, this was observed by
Weinberg et al. (\cite{Weinberg97a}). Varying the mass-resolution by a factor
of eight, had severe effects on the outcome of their results. Navarro
\& Steinmetz (\cite{Navarro97a}) performed similar simulations varying the
mass-resolution of identical runs with a factor of six. They found much
smaller effects, that in addition decreased with redshift. However, the mass-
resolution was {\em much} higher than in Weinberg et al.
(In fact, even the lower resolution runs of Navarro \& Steinmetz had slightly
higher resolution as compared to the ``high resolution'' runs of
Weinberg et al.) For the simulations presented here, the absolute resolution
varies, since it is proportional to the total mass, but for comparison the
$10^{12} {\rm M}_{\odot}$ runs are of comparable resolution to those of
Navarro \& Steinmetz. All in all, this is reassuring but not conclusive
evidence that effects from limited numerical resolution are small.
Ultimately, the tenacity of the underlying assumptions must be judged by
comparisons with observations, and with other models that make different
simplifying assumptions.

\begin{appendix}
\section{Appendix}

\subsection{Initial density fields} \label{app_a}

In hierarchical clustering models, structure forms by the gravitational
growth of small inhomogeneities in the density field.
The fluctuations in the density are conveniently characterised by the
density contrast
\begin{equation}
\contr{\vec{x}} = \frac{\dens{\vec{x}} - \rho_b}{\rho_b}.
\end{equation}
where $\rho_b$ is the mean density of the universe.
In an Einstein-de Sitter universe, a
region with a mean density contrast of zero is marginally bound by gravity.
Regions of space where the density contrast is positive are bound and will
eventually collapse.
At early times $|\delta| << 1$ and gravitational dynamics can be
handled by linear theory. The mathematical formulation for this is laid out
in detail by Bardeen et al. (\cite{BBKS86a}), BBKS henceforth.

To single out sites where objects of mass M are likely to form, the
density contrast is first smoothed on the corresponding length scale
\begin{equation}
\scontr{s}{\vec{x}} = \int\contr{\vec{x'}}W_s(|\vec{x'}-\vec{x}|)d^3x'.
\end{equation}
where
\begin{equation}
W_s(|\vec{x}|) = \exp{\left(-\frac{\vec{x}^2}{2 s^2}\right)}/(2 \pi s^2)^{3/2}.
\end{equation}
s is the smoothing length scale and is given by $s$ $\approx$
$(M/\rho_b)^{1/3}$.
If $\contr{\vec{x}}$ is a Gaussian random field then $\scontr{s}{\vec{x}}$
is also a Gaussian random field.
The power spectrum of $\scontr{s}{\vec{x}}$ is given by $P_s(k) = 
P(k)\exp{(-s^2k^2)}$, where
\begin{equation}
P(k)\equiv \mean{|\contrk{k}|^2},
\end{equation}
and $\contrk{k}$ is the Fourier transform of the density contrast, and
$\mean{...}$ is the mean value.
A peak in the field $\scontr{s}{\vec{x}}$ is a point where the mass inside a
spherical region of size s has a maximum. Such points are plausible sites for
the formation of objects of mass M.

Hoffman et al. (\cite{Hoffman92b}) extended the formalism of BBKS to investigate the 
influence of the background density field on a given smaller scale peak.
By proceeding in a similar manner we consider two Gaussian random fields,
a peak field $\delta^p$, and a background field $\delta^b$, which are
defined by their power spectra
\begin{equation}
P^p(k) =
  \left\{ \begin{array}{l@{\quad \quad}l}
  0 & k < k_{lim} \\
  P_s(k)    & k > k_{lim} \\
  \end{array} \right.
\end{equation}
and
\begin{equation}
P^b(k) =
  \left\{ \begin{array}{l@{\quad \quad}l}
  P_s(k) & k < k_{lim} \\
  0    & k > k_{lim} \\
  \end{array} \right.
\end{equation}
The combined field $\delta^p + \delta^b$ is statistically
identical to $\scontr{s}{\vec{x}}$, the density contrast smoothed on a scale s.
By requiring that $\frac{2 \pi}{k_{lim}} >> s$, extremum points in 
$\scontr{s}{\vec{x}}$ will also be extrema of $\delta^p$.
When smoothing on galactic scales, maxima in the density field are assumed to
be progenitors of galaxies.

Dimensionless field values, $\nu$, are conveniently ex\-pres\-sed in units
of $\sigma$, i.e., $\nu = \delta/\sigma$, where
\begin{equation}
\sigma^2 \equiv \int{P(k)\frac{4 \pi k^2dk}{(2\pi)^3}}
\end{equation}
is the mean square density contrast (BBKS).
Galaxies are believed to form around peaks of height $1 < \nu < 3$,
assuming the density field has been smoothed on an appropriate galactic scale.
Specifying an equivalent value $\nu_b$ for the background field, at the
same point, determines the density field in a larger surrounding of
the peak.
$\nu_s$, $\nu^p$ and $\nu^b$, are related by (Hoffman et al. \cite{Hoffman92b})
\begin{equation}
\nu_s \sigma_s = \nu^p \sigma^p + \nu^b \sigma^b.
\end{equation}
Inside radius s, the statistical properties of the density field
are primarily determined by the small scale peak constraint.
But, further away from the peak, the statistical properties depends
on the value of $\nu^b$.
The main effect of the background field is to change the overall density in
the vicinity of the smaller scale peak.

The approach then, is to set up a
density field $\contr{\vec{x}}$, with the constraint that the corresponding
field $\scontr{s}{\vec{x}}$ sho\-uld have a maximum at the center of the
simulated region.
The small scale field $\delta_s$ is constrained by
\begin{equation}
\delta_s(0) = \nu^p \sigma^p.
\end{equation}
To ensure that
the field has an extremum at the center, it is also required to have vanishing
first derivatives there. These constraints specifies the statistical
properties of the field. The second derivative of the field could also
have been used as a constraint, to ensure that the central extremum is in
fact a maximum. This is neglected since it would be a largely redundant
constraint. It is very unlikely that a $\nu^p \approx 2$ extremum,
as is the case here, is a minimum,
and therefore the statistical properties of the field are only marginally
affected by adding the contrary as a constraint.

Hoffmann \& Ribak (\cite{Hoffman92a}) showed how to create a random realization of a Gaussian
field, that is subject to constraints that are linear in the field.
The method is computationally efficient, and it is easy to handle several
constraints. Any quantities, or combinations of, that can be represented within
linear theory, can be used as constraints for the field. In addition to the
``seeding' peak itself, we have found it useful to also specify $v(0) = 0$,
i.e. requiring that the ``peak'' is stationary. This increases the likelihood
that the forming object stays in the center of the simulation volume.

The field is set up in a cubic region of size $L_{cube}$, where
$L_{cube} >> s$. The largest wavelengths that can be represented is
therefore $2\pi/L_{cube}$. The values used for s are shown in Table
\ref{simpars}. The smoothed density field is then split into $\delta^b$
and $\delta^p$, in such a way that $\delta^p$ contains all
wavelengths that can be represented in the cube, and $\delta^b$ contains
all longer wavelengths.

The background field cannot be accurately represented inside the cubic
region, where the density field is constructed, since it consists of
modes with wavelengths exceeding the size of the region. 
These long
wavelength mod\-es can still have important dynamical effects, because,
by changing the overall density inside the cube, these modes change
the collapse times for
structures in the region. A constant density contrast of $\delta_{b} = 
\nu^b \sigma^b$ was added to the region
to approximate the lowest order effects of the background field.
All wavelengths longer than the expansion volume will be assumed to
give a constant shift of the density.

There is, however, by no means a one-to-one
correspondence between such density peaks and the galaxies that later form
(Katz et al. \cite{Katz93a}).
Objects that form when peaks collapse at a high redshift can at later times
merge with other collapsed objects, and form new, more massive, objects.
Peaks in the density field at a high redshift will then not be in a one to one
correspondence with the galaxies that later form.
This is a nonlinear dynamical process that cannot be handled by linear
theory.
N-body calculations have shown that a significant number
of the galactic halos that form, do in fact correspond to peaks in
the early density field. It may be argued that the ``peaks formalism''
would generate atypical initial conditions. This should not be the case
however, as long as the parameters are within statistically reasonable limits,
i.e. $\nu \leq 3 - 4$. The generated field would still be a statistically
typical one.

The gravitational particle smoothing used in the simulations are shown
in Table \ref{simpars}.


\begin{table}
\caption[]{\label{simpars} Simulation parameters.}
\begin{flushleft}
\begin{tabular}{cllllllllllllll}
\hline\noalign{\smallskip}
Total mass & $r_{init}$ & $\nu^p$ & $\nu^b$ & $L_{cube}$ \\
\noalign{\smallskip}
[${\rm M}_{\odot}$] & [Mpc] & & & [Mpc] & \\
\noalign{\smallskip}
\hline\noalign{\smallskip}
$10^{9}$ & 0.15 & 2.0 & 2.0 & 2.0 \\
$10^{10}$ & 0.33 & 2.0 & 2.0 & 4.0 \\
$10^{11}$ & 0.71 & 2.0 & 2.0 & 8.0 \\
$10^{12}$ & 1.52 & 2.0 & 2.0 & 16.0 \\
$10^{13}$ & 3.27 & 3.0 & 3.0 & 32.0 \\
\noalign{\smallskip}
\hline
\end{tabular}
\end{flushleft}
\end{table}



\subsection{Two-body heating} \label{two_body}

\begin{figure}
  \psfig{file=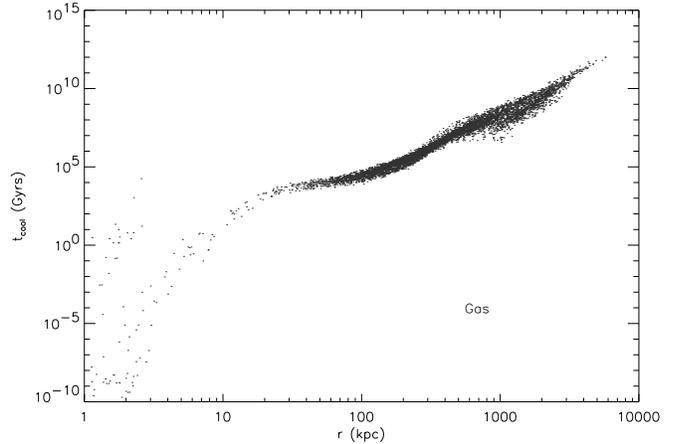,width=8.8cm}
  \caption{ \label{cool_time} Cooling times at $z = 0$, for the
$10^{13} {\rm M}_{\odot}$ run with metallicity. This simulation has the shortest
cooling times, making potential two-body heating effects easier to detect.
(There's equally long or longer cooling times in the $10^{12} {\rm M}_{\odot}$ run. Some particles in the center, have longer cooling times. This is because
they have temperatures below the ``drop'', $T \sim 10^4$ K, in the cooling
function, and therefore have essentially no cooling.)}
\end{figure}

\begin{figure}
  \psfig{file=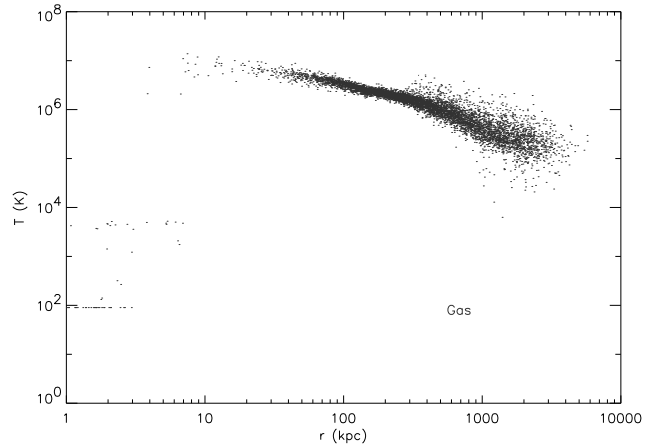,width=8.8cm}
  \caption{ \label{temp_rad} Temperature at $z = 0$, for the
$10^{13} {\rm M}_{\odot}$ run with metallicity.}
\end{figure}

There have been recent results showing that there can be undesired, artificial
heating of the gas component, due to two-body relaxation like effects,
caused by the
dark matter halo (Steinmetz \& White \cite{Steinmetz97a}). This could
be the case for our simulations, where the dark matter particles are of
higher mass. (i.e. the simulations of $10^{12}$ and $10^{13} {\rm M}_{\odot}$
total mass.) The effect would be that, e.g., the cooling flow might be reduced.
In the hot gas halo, the cooling times are on the other hand very long,
as can be seen from Fig. \ref{cool_time}, and it cannot cool within a Hubble
time. Corresponding temperatures are shown in Fig. \ref{temp_rad}. Only in the
absolute center, (i.e. the disk), the cooling times are short. A few
particles just within 10 kpc seems to be possibly prevented from cooling,
so we cannot rule out that there are some two-body heating effects. We doubt,
however, that a large fraction hot gas halo, is being held at virial
temperatures, (or was formed), solely due to two-body effects. However, we
believe these two-body heating effects should be taken seriously, and
avoided in future work.

\end{appendix}

\begin{acknowledgements}
We are grateful to the theoretical astrophysics group, at Uppsala Astronomical
observatory, in particular for providing us with the computational resources
needed.  
\end{acknowledgements}

\def\apj{ApJ}
\def\apjs{ApJS}
\def\mnras{MNRAS, }
\def\aap{A\&A}

\end{document}